\documentstyle[prb,eqsecnum,preprint,tighten,aps]{revtex}
\begin{document}
\draft
\title{Casimir Energy of a Semi-Circular Infinite Cylinder}
\author{V.V.~Nesterenko\thanks{Electronic address: nestr@thsun1.jinr.ru}}
\address{Bogoliubov Laboratory of Theoretical Physics,
Joint Institute for Nuclear Research, 141980 Dubna,  Russia.}
\author{G.~Lambiase\thanks{Electronic address: lambiase@sa.infn.it} and G. Scarpetta
\thanks{Electronic address: lambiase@sa.infn.it}}
\address{Dipartimento di Scienze Fisiche ``E.R. Caianiello'',
 Universit\'a di Salerno, 84081 \\ Baronissi (SA), Italy.\\
 INFN, Sezione di Napoli, 80126 Napoli, Italy.}
\date{\today}
\maketitle
\begin{abstract}
The Casimir energy of a semi-circular  cylindrical shell is
calculated by making use of the zeta function technique.
This shell is obtained by crossing an infinite
circular cylindrical shell by a plane passing through
the symmetry axes of the cylinder and by considering only a half of
this configuration.  All the surfaces, including the cutting plane,
are assumed to be perfectly conducting. The zeta
functions for scalar massless fields obeying the Dirichlet and
Neumann boundary conditions on the semi-circular cylinder are
constructed exactly. The sum of these zeta functions gives the zeta
function for electromagnetic field in question. The relevant plane
problem is considered also.  In all the cases the final expressions
for the corresponding Casimir energies contain the pole
contributions which are the consequence
of the edges or corners in the boundaries. This implies that further renormalization is needed in
order for the finite physical values for vacuum energy to be
obtained for given boundary conditions.  \end{abstract}

\section{Introduction}
     When calculating the ground state energy of a quantum field (the
Casimir energy) the main problem is to single out a finite part of
the  vacuum energy which is initially divergent. Usually for this
purpose a subtraction procedure  is used with preliminary
regularization of the divergent expressions (for example, by
introducing ultraviolet cutoff). However in quantum field theory
treated with allowance for nontrivial boundary conditions or in the
space-time with curvature, a complete renormalization procedure is
not formulated explicitly. Therefore, for any specific problem the
subtraction procedure should be invented anew. As a result, one
succeeds in calculating the Casimir energy only in the problems with
known spectra or at least implicitly known spectra. Practically it
implies the boundary conditions of high symmetry~\cite{PRep,MT}
(parallel plates, sphere, cylinder).

     In studies of the Casimir energy it is wildly used the zeta
function technique~\cite{Od,ten} which is also referred to as the
zeta regularization or zeta renormalization. In fact, the use of the
zeta functions, as well as other regularizations, gives only
regularized quantities for  ground state energy, for effective
potential and so on. The necessity to renormalize the expressions
obtained in this way certainly remains.  However, in some problems
the zeta technique gives at once a finite result. Usually the latter
is considered to be a renormalized physical answer though generally
it is not the case.\cite{Bordag-new}

     When using the zeta regularization in one or another problem, it
is desirable to know beforehand whether the finite result can be
obtained in this way. In order to answer this question the general
analysis of the divergences in the problem at hand should be
accomplished. This can  be done by calculating the heat kernel
coefficients\cite{Bor} depending on the geometry of the manifold
under consideration. For a large class of situations these
coefficients have been obtained.\cite{Branson} However there is a
number of problems (for example, boundaries with edges or corners)
for which no general results regarding the heat trace are known.

   In this situation it is undoubtedly worth carrying out, in the
framework of the zeta function technique, the calculations of the Casimir
energy for new configurations, both the cases being interesting
with finite result and with pole contributions left in the final
expression for the vacuum energy.

      In the present paper we address the calculation of
the Casimir energy for boundaries with edges, more precisely,
the vacuum energy of electromagnetic field will be calculated for a
semi-circular cylindrical shell by making use of the relevant
zeta functions. This shell is obtained by crossing an infinite
circular cylindrical shell by a plane passing through
the symmetry axes of the cylinder. All the surfaces, including the
infinite cutting plane, are assumed to be perfectly conducting.
Obviously it is sufficient to consider only a half of this
configuration (left or right) which we shall refer to as a
semi-circular cylindrical shell or, for sake of shortening, as a
semi-circular cylinder. The internal boundary value problem for this
configuration is nothing else as a semi-cylindrical waveguide. In the
theory of waveguides\cite{HdP1} it is well known that a semi-circular
waveguide has the same eigenfrequencies as the cylindrical one but
without degeneracy (without doubling) and safe for one frequency
series (see below).  Notwithstanding the very close spectra, the zeta
function technique does not give a finite  result for a semi-circular
cylinder unlike for a circular one. First the Casimir energy of an infinite
perfectly conducting cylindrical shell has  been calculated in Ref.\
\onlinecite{DRM} by introducing ultraviolet cutoff and recently
this result was derived by zeta function technique\cite{MNN} (see also Refs.
\onlinecite{LNB,NPcyl,GRom}). As far as we know the asymmetric
boundaries such as  a semi-circular cylinder have not been considered
in the Casimir problem.

The paper is organized as follows. In Sec.\ II the electromagnetic
spectra are considered in details for cylindrical and
semi-cylindrical shells. The general solution of the Maxwell
equations for boundary conditions chosen is expressed  in terms of two scalar
functions, longitudinal components of the electric and magnetic
Hertz vectors. These scalar functions are the eigenfunctions of
the two-dimensional transverse Laplace operator and obey  the
Dirichlet and Neumann boundary conditions on the conducting
surfaces. In Sec.\ III the spectral zeta function is constructed
for the Dirichlet boundary value problem. To this end, the
technique is used which has been elaborated before for
representing the spectral zeta function, with given eigenfrequency
equations, in terms of contour integral. When carrying out the
analytic continuation of the zeta function into the physical
region, the uniform asymptotic expansion for the modified Bessel
functions is used. In the same way, in Sec.\ IV the zeta function
is constructed for a scalar field obeying the Neumann boundary
conditions given on the surface of a semi-circular cylindrical shell.
The Section V is concerned with the complete zeta function for
electromagnetic field with boundary conditions on the
semi-circular cylinder. Transition to the relevant
two-dimensional problem is also considered here. In the
Conclusion (Sec.\ VI) the results obtained are summarized, and the
origin of the pole singularities of the zeta functions at hand and
their relation to the respective boundary value problem are
briefly discussed.

\section{Eigenmodes of Electromagnetic Field for  Circular and
Semi-Circular Cylinders}
The construction of the solutions to the Maxwell equations with
boundary conditions given on closed surfaces proves to be
nontrivial problem. Mainly it is due to the vector character of the
electromagnetic field \cite{HdP1,Str,HdP2}. In the case of
cylindrical symmetry the electric ${\bf E}$ and magnetic ${\bf H}$
fields are expressed in terms of the electric (${\bf \Pi}^{\prime}$)
and magnetic (${\bf \Pi}^{\prime\prime}$) Hertz vectors having only
one non-zero component
\begin{eqnarray}\label{P1}
 {\bf \Pi}^{\prime} & = & {\bf e}_z \Phi (r, \varphi)\,e^{\pm i k_z'
 z} \,,\\
 {\bf \Pi}^{\prime\prime} & = & {\bf e}_z \Psi (r, \varphi)\,e^{\pm i
 k_z'' z} \label{P2}\,.
\end{eqnarray}
Here the cylindrical coordinate system $r, \varphi, z$ is used
with $z$ axes directed along the cylinder axes. The common
time--dependent factor $e^{i\omega t}$ is dropped. The scalar
functions $\Phi (r, \varphi)$ and $\Psi (r, \varphi)$ are the
eigenfunctions of the two--dimensional transverse Laplace operator
and meet, respectively, the Dirichlet and Neumann conditions on
the boundary $\partial\Gamma$
\begin{equation}\label{D}
  ({\bbox \nabla}^2_{\bot}+\gamma^{'\,2})\Phi(r, \varphi)=0\,, \qquad
  \left. \Phi(r, \varphi)\right|_{\partial\Gamma}=0 \,,
\end{equation}
\begin{equation}\label{N}
  ({\bbox \nabla}^2_{\bot}+\gamma''^{2})\Psi(r, \varphi)=0\,, \qquad
 \left. \frac{\partial\Psi(r, \varphi)}{\partial n}\right|_{\partial\Gamma}
=0 \,,
\end{equation}
where ${\bbox \nabla}^2_{\bot}$ is the transverse part of the
Laplace operator
\begin{equation}\label{Ltr}
  {\bbox \nabla}^2_{\bot}=\frac{\partial^2}{\partial r^2}+\frac{1}{r}\frac{\partial}{\partial r}
  +\frac{1}{r^2}\frac{\partial^2}{\partial\varphi^2}
\end{equation}
and
\begin{equation}\label{gamma}
  \gamma^{'\, 2}=\omega^2-k_z^{'\, 2}\,, \qquad
  \gamma^{''\, 2}=\omega^2-k_z^{''\, 2}\,.
\end{equation}

First we consider a cylindrical shell. In this case the functions
$\Phi (r, \varphi)$ and $\Psi (r, \varphi)$ should be
$2\pi$-periodic in angular variable $\varphi$. As a result the
Dirichlet boundary value problem (\ref{D}) has the following
unnormalized eigenfunctions ($E$-modes)
\begin{equation}
  \Phi_{nm}(r, \varphi) =   {\sin \atop \cos} (n\varphi)
\cases{J_n(\gamma_{nm}' r), \quad r<a ,\cr
H_n^{(1)}({\bar \gamma}_{nm}'\, r), \quad
                        r>a, \cr} \label{efd}
\end{equation}
where $a$ is the cylinder radius, $J_n(x)$ are the Bessel functions,
$H^{(1)}(x)$ are the Hankel functions of the first kind,
and $\gamma'_{nm}$, $\bar{\gamma}_{nm}'$ stand for the roots of the
frequency equations
 \begin{eqnarray}
\label{fed}
   &J_n(\gamma_{nm}'\, a)=0, \quad
 H_n^{(1)}(\overline{\gamma}_{nm}'\, a)=0,&  \\ &n=0, 1,
 2, \ldots , \quad m=1, 2, \ldots \,.& \nonumber \end{eqnarray}

For the Neumann boundary value problem (\ref{N}) we have the
$H$-modes
\begin{equation}
  \Psi_{nm}(r, \varphi) ={\sin \atop \cos} (n\varphi)
  \cases{J_n(\gamma_{nm}'' r), \quad r<a, \cr
H_n^{(1)}(\bar{\gamma}{''}_{nm}\, r),
 \quad r>a, \cr} \label{efn}
\end{equation}
where $\gamma''_{nm}$ and  $\overline{\gamma}_{nm}''$ are the roots of the
equations
 \begin{eqnarray}
   &\left. \frac{\displaystyle d}{\displaystyle
dr}J_n(\gamma''_{nm}\, r)\right|_{r=a}=0, \quad
   \left. \frac{\displaystyle d}{\displaystyle
dr}H^{(1)}(\overline{\gamma}_{nm}^{\,''}\, r)\right|_{r=a}=0\,,&
\label{fen} \\
 &n=0, 1, 2, \ldots , \quad m=1, 2, \ldots \,. & \nonumber
\end{eqnarray}
As usual, it is assumed that for $r>a$ the eigenfunctions should
satisfy the radiation condition.

It is important to note that each root
\begin{equation}\label{dr}
 \gamma'_{nm}, \quad \overline{\gamma}_{nm}^{\,'}, \quad
 {\gamma}_{nm}^{\,''}, \quad \overline{\gamma}_{nm}^{\,''}, \quad
 n\geq 1, \quad m\geq 1
\end{equation}
is doubly degenerate since, according to Eqs.\ (\ref{efd}),
(\ref{efn}), there are two eigenfunctions which are
proportional to either $\sin (n\varphi)$ or $\cos (n \varphi)$.
The frequencies with $n=0$
\begin{equation}\label{ndr}
 \gamma'_{0m}, \quad \overline{\gamma}_{0m}^{\,'}, \quad
 {\gamma}_{0m}^{\,''}, \quad \overline{\gamma}_{0m}^{\,''}, \quad
 m= 1, 2, \ldots
\end{equation}
are independent on $\varphi$, and the degeneracy disappears.

For given Hertz vectors ${\bf \Pi}'$ and ${\bf \Pi}''$ the electric
and magnetic fields are constructed by the formulas
 \begin{eqnarray}
 {\bf E}&= &{\bbox \nabla}\times {\bbox \nabla} \times {\bf \Pi}'\,,
 \quad {\bf H}=-i\omega {\bbox \nabla}\times {\bf \Pi}' \qquad
 (E\text{-modes})\,,\ \nonumber \\
 \label{fields}
 {\bf E}& =& i\omega {\bbox \nabla} \times {\bf \Pi}''\,,
 \qquad {\bf H}={\bbox \nabla}\times {\bbox \nabla}\times
 {\bf \Pi}'' \qquad
 (H\text{-modes})\,.
\end{eqnarray}
It has been proved \cite{Heyn} that the superposition of these modes
gives the general solution to the Maxwell equations in the problem
under consideration. An essential merit of using the Hertz
polarization vectors is that in this approach the necessity to
satisfy the gauge conditions does not arise.

Let us consider a waveguide which is obtained by cutting the
infinite cylindrical shell by a plane passing through the symmetry
axes of the cylinder (see Fig.~1). All the surfaces are assumed
to be perfectly conducting.
In this case the boundary value problems
(\ref{D}) and (\ref{N}) for the Hertz electric (${\bf \Pi}'$) and
magnetic (${\bf \Pi}''$) vectors have the following eigenfunctions
\begin{eqnarray}
 & \Phi_{nm}(r, \varphi)=\sin (n\varphi)\cases{
  J_n(\gamma_{nm}'\,r)\,, \quad r<a\,, \cr
         H_n^{(1)}(\overline{\gamma}_{nm}'\,r)\,, \quad r>a \,,
         \cr }& \label{fds}  \\
 &n=1, 2, \ldots \,, \qquad m=1, 2, \ldots \, & \nonumber
\end{eqnarray}
 and
\begin{eqnarray}\label{fns}
  &\Psi_{nm}(r, \varphi)=\cos (n\varphi) \cases{
  J_n(\gamma_{nm}''\,r)\,, \quad r<a\,, \cr
         H_n^{(1)}(\overline{\gamma}_{nm}''\,r)\,, \quad r>a\,,\cr} & \\
   &n=0, 1, 2, \ldots \,, \qquad m=1, 2, \ldots \,. & \nonumber
\end{eqnarray}
The frequencies $\gamma_{nm}'$, $\overline{\gamma}_{nm}'$,
$\gamma_{nm}''$, and $\overline{\gamma}_{nm}''$ are determined by
the same equations (\ref{fed}) and (\ref{fen}). However the new
spectral problem has two essential distinctions: i) the
frequencies (\ref{dr}) are now nondegenerate, and ii) two series
of eigenfrequencies
\begin{equation}\label{absent}
  \gamma_{0m}', \quad \overline{\gamma}_{0m}'\,, \quad m=1, 2,
  \ldots
\end{equation}
are absent. At first sight one could expect that such a change of the
spectrum cannot influence drastically on the ultraviolet behaviour of
the relevant spectral density. However, as it will be
shown below, the zeta function for a semi-circular cylinder, unlike
for a circular one, does not provide a finite answer for the Casimir
energy in the problem in question.

In view of all above-mentioned the zeta function for electromagnetic field
obeying  the boundary conditions on the surface of the semi-circular
cylinder is the sum of two zeta functions for scalar massless  fields
satisfying the Dirichlet and Neumann conditions on the lateral of this cylinder.

\section{Zeta Function for Dirichlet boundary value problem}
First we consider the Dirichlet boundary conditions.
We shall proceed from the following representation for the zeta
function in terms of a contour integral for given frequency
equations (\ref{fed}) with $n=1, 2, \ldots$
\begin{equation}\label{contour}
\zeta_{\text{cyl}}^{\text{D}}(s)=\frac{1}{2\pi i}\int_{-\infty}^{\infty}
 \frac{d
k_z}{2 \pi }\sum\limits_{n=1}^{\infty}
\oint_{C}^{}d \gamma\,(\gamma^2+k_z^2)^{-s/2}\frac{d}{d\gamma}
\ln\frac{J_n(\gamma a) H_n^{(1)}(\gamma a)}{J_n(\infty)
H_n^{(1)}(\infty)}.
\end{equation}
The contour $C$ consists of the
imaginary axis $(-i\infty,i\infty)$ and a semi-circle of an infinite
radius in the right half-plane of a complex variable $\gamma$. The
details of obtaining this integral representation can be found in
Refs.~\onlinecite{Bordag,MNN,LNB,NP}. Contribution into Eq.\
(\ref{contour}) of integration along a semi-circle of infinite radius
vanishes.  Therefore upon integration over $k_z$ this formula
acquires the form
\begin{equation}
\label{3.2}
\zeta^{\text{D}}(s)=C(s)\sum\limits_{n=1}^{\infty}\int_{0}^{\infty} dy\,
y^{1-s}\frac{d}{dy}\ln\left[2y I_n(y) K_n(y)\right]
\end{equation}
with
\begin{equation}
\label{3.3}
C(s)=\frac{a^{s-1}}{2\sqrt{\pi} \Gamma\left(\frac{\displaystyle s}
{\displaystyle 2}\right)\Gamma\left(\frac{\displaystyle 3-s}
{\displaystyle 2}\right)}\,{.}
\label{3.2a}
\end{equation}

     In order to accomplish the analytic continuation of~(\ref{3.2})
into the physical region including the point $s=-1$, we shall use the
uniform asymptotic expansion for the modified Bessel functions\cite{AS}
\begin{eqnarray}\label{3.4}
\ln\left[2 y n I_n(ny)K_n(ny)\right]&=& \ln(y
t)+\frac{t^2}{8 n^2}(1-6t^2+5t^4)\nonumber\\ &&
+\frac{t^4}{64 n^4}(13-284 t^2+ 1062 t^4-1356 t^6+ 565 t^8) +O(n^{-6})\,{,}
\end{eqnarray}
where $t=1/\sqrt{1+y^2}$. Following the usual procedure applied in the
analogous calculations,\cite{Milton1,Milton,LR,CEK} we add and
subtract in the integrand in Eq.\ (\ref{3.2}) the first two terms of
the asymptotic expansion (\ref{3.3}). After that we combine all the
terms there in the following way
\begin{eqnarray}
 \zeta^{\text{D}}_{\text{cyl}}(s)&=&C(s)\left[Z_1(s)+Z_2(s)+Z_3(s)\right],
  \label{3.5}\\
 Z_1(s)&=&\frac{1}{2}\sum\limits_{n=1}^{\infty}n^{1-s}\int_{0}^{\infty} dy\,
    y^{1-s}\frac{d}{dy}\ln \left(\frac{y^2}{1+y^2}\right),
      \label{3.6}\\
 Z_2(s)&=&\frac{1}{8}\sum\limits_{n=1}^{\infty}n^{-1-s}
   \int_{0}^{\infty} dy\,
   y^{1-s}\frac{d}{dy}\left[t^2(1-6\, t^2+5\, t^4)\right], \label{3.7}\\
Z_3(s)&=&\sum_{n=1}^{\infty}n^{1-s}
\int_0^{\infty} dy\, y^{1-s}\frac{d}{dy}
\biggl [\ln(2yn \,I_n(yn) K_n(ny)) \nonumber \\
&&-\left .\ln\frac{y}{\sqrt{1+y^2}}-\frac{t^2(1-6 t^2+5 t^4)}{8 n^2}\right]{.}
\label{3.8}
\end{eqnarray}

Analytic continuation of the function $Z_1(s)$
into vicinity of the point $s=-1$ can be accomplished in the same
way as it has been done in Ref.\ \onlinecite{NP}. Therefore we write here
only the final result of this continuation
\begin{equation}
Z_1(s)=\frac{1}{2} \zeta(s-1)\Gamma\left(\frac{3-s}{2}\right)
\sum\limits_{m=1}^{\infty}\frac{\displaystyle \Gamma\left(m-\frac{\displaystyle
1-s}{\displaystyle 2}\right)}{m \Gamma(m)}.
  \label{3.9}
\end{equation}

 The integral in Eq.\ (\ref{3.6})
converges when $-1<\text{Re } s< 3$, and the sum over
$n$ is finite for Re$\, s>0$. Thus, the
regions, where the integral and the sum exist, overlap, and this
formula can be used for constructing the analytic continuation
needed. For this aim we substitute the sum by the Riemann zeta
function
\begin{equation}\label{3.10}
  \sum\limits_{n=1}^{\infty}n^{-1-s}=\zeta (s+1)
\end{equation}
and define the integral as an analytic function by making use
of the formula\cite{GR}
\begin{equation}\label{3.11}
\int_0^{\infty}dy \,y^{1-s}\frac{d}{dy}t^{2(\rho-1)}=
(1-\rho)\,\frac{\Gamma\left(\frac{\displaystyle 3-s}{\displaystyle
 2}\right)
\Gamma\left(\rho-\frac{\displaystyle  3-s}{\displaystyle  2}
\right)}{\Gamma(\rho)}, \quad 3-2\mbox{ Re }\rho<\mbox{ Re }s<3.
\end{equation}
In view of the poles of the gamma functions on the right-hand side
of this relation, the integral on the left-hand side of it is
well defined, as a function of the complex variable $s$, only in
the region indicated in Eq.\ (\ref{3.11}). Doing the analytic
continuation of this integral we define it outside this region
also by this equation, keeping in mind that the gamma functions
involved should be treated as the analytic functions over all the
plane of the complex variable $s$ safe for the known poles. This
gives
\begin{equation}
\label{3.12}
Z_2(s)=\frac{1}{8}\zeta(s+1)
\Gamma\left(\frac{3-s}{2}\right)\Gamma\left(\frac{1+s}{2}\right)
\left[-1+3(1+s)-\frac{5}{8}(3+s)(1+s)\right].
\end{equation}

In order to investigate the convergence of the integral entering
in Eq.\ (\ref{3.7}) it makes sense to substitute in the integrand the
logarithmic function by expansion (\ref{3.3}). After that it is easy
to be convinced that the integral under consideration converges when
$-3<\text{Re } s<3$. The sum over $n$ in this formula is finite for
$\text{Re } s>-2$. Hence, the function $Z_3(s)$ is an analytic
function without singularities in the domain $-2<\text{Re }s<3$. It
is quiet enough for our purpose, and the analytic continuation is
unnecessary.

  Summarizing we conclude that Eqs.\ (\ref{3.3}), (\ref{3.5}),
(\ref{3.8}), (\ref{3.9}), and (\ref{3.12}) afford the analytic
continuation needed and define the zeta function $\zeta
^{\text{D}}(s)$ as  an analytic function in the region including the
point $s=-1$.


Now we are able to calculate the value of the zeta function
$\zeta^{\text{D}}(s)$ at the point $s=-1$. For the coefficient $C(s)$ in
Eq.\ (\ref{3.3}) we have
\begin{equation}\label{3.13}
  C(-1)=-\frac{1}{4\pi a^2}\,.
\end{equation}
From Eq.\ (\ref{3.9}) it follows that
\begin{equation}\label{3.14}
Z_1(-1)=\frac{1}{2}\lim_{s\to -1}\zeta(s-1)\left[
\Gamma\left(\frac{1+s}{2}\right)
+\sum\limits_{m=2}^{\infty}\frac{1}{m(m-1)}\right].
\end{equation}
With allowance for the relations
\begin{equation}\label{3.15}
\Gamma(x)=\frac{1}{x}-\gamma+O(x),\quad
\sum\limits_{m=2}^{\infty}\frac{1}{m(m-1)}=1,\quad \zeta(-2)=0,
\end{equation}
where $\gamma $ is the Euler constant, $\gamma =0.577215\ldots$,
one derives
\begin{eqnarray}
Z_1(-1)&=&\lim_{s\to-1}^{}\frac{1}{2}\left[\zeta(-2)+\zeta'(-2)(s+1)
 +O\left((s+1)^2\right)\right]\left[\frac{2}{s+1}-\gamma+O(s+1)\right]\nonumber
 \\
 &=&  \zeta'(-2)=-0.030448\,{.}
 \label{3.16}
\end{eqnarray}
Using the values of the Riemann zeta function and its derivative at the
origin
 \[
 \zeta(0)=-\frac{1}{2},\quad
\zeta'(0)=-\frac{1}{2}\ln(2\pi)
 \]
and taking into account the behaviour of the gamma function near
zero (see Eq.\ (\ref{3.15})) we deduce from Eq.\ (\ref{3.12})
\begin{eqnarray}
Z_2(-1)&=&\frac{1}{8}\lim_{s\to-1}^{}\left[\zeta(0)+\zeta'(0)(s+1)
+O\left((s+1)^2\right)\right] \nonumber \\
  & &\times \left[\frac{2}{s+1}-\gamma+{\cal O}\left( s+1 \right)
\right]\cdot
 \left[-1+\frac{7}{4}(s+1)\right]  \nonumber\\
 &=&-\frac{7}{32}-\frac{\gamma}{16}+\frac{1}{8}\ln(2\pi)+\left.
 \frac{1}{8}\frac{1}{s+1}\right|_{s\to -1}{.} \label{3.17}
\end{eqnarray}

When calculating $Z_3(-1)$ we shall use Eq.\ (\ref{3.8}) for sevral first
values of $n,\quad n\le n_0$ and for $n>n_0$ we substitute the asymptotic
expansion (\ref{3.4}) into (\ref{3.8}) with the result
\begin{eqnarray}
Z_3^{\text{as}}(s)
&=&\frac{1}{64}\left ( \sum\limits_{n=n_0+1}^{\infty}n^{-3-s}\right )
      \int_{0}^{\infty} dy\, y^{1-s}\frac{d}{dy}\left[t^4(13-284\,
      t^2+1062\, t^4-1356\, t^6 +565\, t^8)\right] \nonumber \\
&=&
\frac{1}{64}\left(\sum\limits_{n=n_0+1}^{\infty}n^{-3-s}\right )
\Gamma\left(\frac{3-s}{2}\right)
\left[-13\, \Gamma\left(\frac{3+s}{2}\right)+ 142\,
\Gamma\left(\frac{5+s}{2}\right)-\frac{532}{3}\,
\Gamma\left(\frac{7+s}{2}\right)\right.\nonumber\\
&&\left.+\frac{113}{2}\,
\Gamma\left(\frac{9+s}{2}\right)-\frac{113}{24}\,
\Gamma\left(\frac{11+s}{2}\right)\right].
\label{z3as}
\end{eqnarray}
The value of $n_0$ should be chosen so as to provide the accuracy
needed. This algorithm with $n_0=6$ gives for $Z_3(-1)$
\begin{equation}\label{3.18}
  Z_3(-1)= 0.022806
\end{equation}
Summing up Eqs.\ (\ref{3.16}), (\ref{3.17}), and (\ref{3.18}) we
obtain
\begin{eqnarray}
 \zeta^{\text{D}}(-1) & = & -\frac{1}{4\pi
 a^2}\left(-\frac{7}{32}+ 0.022806 -\frac{\gamma}{16}+\frac{1}{8}\ln
 (2\pi)+\zeta'(-2)+\left.
 \frac{1}{8}\frac{1}{s+1}\right|_{s\to -1}\right) \nonumber \\
 & = & \frac{1}{a^2}\left(0.000523 - 0.009947 \left.
\frac{1}{s+1}\right|_{s\to -1}
 \right){.}
\label{3.19}
\end{eqnarray}
Thus the zeta function $\zeta^{\text{D}}(s)$ has a pole at the point
$s=-1$, therefore it does not give the finite (renormalized) value
for the  respective Casimir energy
\begin{equation}\label{3.20}
  E^{\text{D}}=\frac{1}{2}\, \zeta^{\text{D}}(-1)\,.
\end{equation}
It implies that further renormalization is required.
\section{Zeta Function for Neumann Boundary Value Problem}
When constructing the zeta function for the
boundary value problem (\ref{N}) with $\partial \Gamma$ being a
semi-circular infinite cylinder, we shall again proceed from the
frequency equations (now from Eq.\ (\ref{fen}). It should be taken
into account that all these roots are not degenerate. Therefore
we can write analogously to Eq.\ (\ref{contour})
\begin{equation}\label{4.1}
\zeta^{\text{N}}(s)=\frac{1}{2\pi i}\int_{-\infty}^{\infty} \frac{d
k_z}{2 \pi }\sum\limits_{n=0}^{\infty}
\oint_{C}^{}d\gamma\; (\gamma^2+k_z^2)^{-s/2} \frac{d}{d\gamma}
\ln\frac{J_n'(\gamma
a) H_n^{(1)\, '}(\gamma a)}{J_n'(\infty) H_n^{(1)\, '}(\infty)}.
\end{equation}
The contour $C$ is the same as in Eq.\ (\ref{contour}) and the
prime on the Bessel and Hankel functions denotes differentiation
with respect to the entire argument.

The product of the derivatives of the modified Bessel functions
$I_n'(z)K_n'(z)$ has
the following asymptotics when $n$ is fixed and $\vert z\vert$ is
large \cite{AS}
\begin{equation}\label{4.2}
  I_n'(z)K_n'(z)=-\frac{1}{2z}\left[1+\frac{4n^2-3}{2
  (2z)^2}+\frac{(4n^2-1)(4n^2-45)}{8(2z)^4}+O(z^{-6})\right]\,.
\end{equation}
Taking this into account in calculation of the denominator in Eq.\
(\ref{4.1}), we obtain for $\zeta^{\text{N}}(s)$ upon integration over
$k_z$
\begin{equation}\label{4.3}
\zeta^{\text{N}}(s)=C(s)\sum\limits_{n=0}^{\infty}\int_{0}^{\infty}
dy\, y^{1-s}\frac{d}{dy}\ln\left[-2y I_n'(y) K_n'(y)\right]
\end{equation}
with the same function $C(s)$ as in Eq.\ (\ref{3.3}).

 Further we shall use
the uniform asymptotic expansion
for the derivatives of the  Bessel functions\cite{AS}
\begin{eqnarray} \ln\left[-2 y n I_n'(ny)K_n'(ny)\right]&=& -\ln(y
t)+\frac{t^2}{8 n^2}
(-3+10t^2-7t^4)+\frac{t^4}{n^4}\left(-\frac{27}{64} \right.\nonumber \\
   & &\left. +\frac{109}{16}
t^2-\frac{733}{32} t^4+\frac{441}{16} t^6-\frac{707}{64}
t^8\right) +{\cal O}(n^{-6})\,{.} \label{4.4}
\end{eqnarray}
In order to render  the integral in the term with $n=0$ in Eq.\ (\ref{4.3})
convergent  we
add and subtract  the second term from the
asymptotics (\ref{4.4}). For $n\ge 1 $ in Eq.\
(\ref{4.3}) we add and subtract in respective intagrands the first
two terms of the asymptotic expansion (\ref{4.4}). After that we
 combine all the terms in the following way
\begin{eqnarray}
\zeta^{\text{N}}(s)&=&C(s)\left[ V_0(s)+V_1(s)+V_2(s)+V_3(s)\right],
   \label{4.5} \\
 V_0(s)&=&\int_0^{\infty} dy\, y^{1-s}\frac{d}{dy} \left\{\ln[-2y
          I_0'(y) K_0'(y)]-\frac{t^2}{8}(-3+10
            t^2-7t^4)\right\},\label{4.6} \\
 V_1(s)&=&-\frac{1}{2}\sum\limits_{n=1}^{\infty}n^{1-s}\int_{0}^{\infty} dy\,
           y^{1-s}\frac{d}{dy}\ln \left(\frac{y^2}{1+y^2}\right)
            =-Z_1(s)\,{,} \label{4.7} \\
 V_2(s)&=&\frac{1}{8}\left(\sum\limits_{n=1}^{\infty}n^{-1-s}+
         1 \right)\int_{0}^{\infty} dy\, y^{1-s}\frac{d}{dy}\left[t^2(-3+10\,
         t^2-7\, t^4)\right], \label{4.8} \\
 V_3(s)&=&\sum\limits_{n=1}^{\infty}n^{1-s}
         \int_{0}^{\infty} dy\, y^{1-s}\frac{d}{dy}\biggl \{\ln \left [
-2ynI_n'(ny)K_n'(ny)
\right ] \nonumber \\
&& \left .
+\ln (yt) -\frac{t^2}{8n^2}(-3+10t^2-7t^4)
\right \}{.}               \label{4.9}
\end{eqnarray}
Taking into account the behaviour of the product $I'_0(y)K'_0(y)$
at the origin and at infinity
 \[
 -2y I_0'(y)K_0'(y)=y+\frac{1}{8}(-1+4y-4\ln 2+\ln y)y^3+O(y^5\ln
 y)\,,
 \]
\begin{equation}\label{4.10}
  -2y
  I_0'(y)K_0'(y)=1-\frac{3}{8y^2}+\frac{45}{128y^4}+O(y^{-6})
\end{equation}
it is easy to show that Eq.\ (\ref{4.6}) defines $V_0(s)$ as an
analytic function in the region $-3<\text{Re }s<1$. Under this
condition the integration by parts can be done here
\begin{equation}\label{4.11}
 V_0(s)=-(1-s)\int_0^{\infty} dy\, y^{-s}\left\{\ln[-2y
          I_0'(y) K_0'(y)]-\frac{t^2}{8}(-3+10
            t^2-7t^4)\right\}{.}
\end{equation}

The function $V_1(s)$ differs only in sign of the function
$Z_1(s)$ from the proceeding Section. The integral in Eq.\
(\ref{4.7}) is convergent when $-1<\text{Re }s<3$. The sum over $n$ in
this formula is finite when Re $s>0$.
Thus the regions, where
the integral and the sum exist, overlap and this formula can be used for
constructing the analytic continuation needed by making use of the
substitutions (\ref{3.10}) and (\ref{3.11}).
Substituting the sum in Eq.\ (\ref{4.8}) by the Riemann zeta
function and doing the integration according to Eq.\ (\ref{3.11}) one obtains
\begin{equation}\label{4.12}
  V_2(s)=\frac{1}{8}[\zeta (1+s)+1]\Gamma\left(\displaystyle{\frac{3-s}{2}}
\right)
  \Gamma\left(\displaystyle{\frac{1+s}{2}}\right)\left[3-5(1+s)
+\frac{7}{8}(1+s)(3+s)
 \right]\,.
\end{equation}

The convergence of the integral in Eq.\ (\ref{4.9}) can be determined
in the same line as it has been done for the function $Z_3(s)$ in
the preceding Section. This  integral converges when
$-3<\text{Re }s<3$, and the sum encountered here is finite for
$\text{Re }s>-2$.  Hence there is no need to do  analytic
continuation for $V_3(s)$.

Finally the zeta function $\zeta^{\text{N}}(s)$ for the massless scalar
field obeying  the Neumann boundary conditions on a semi-circular
cylinder is determined explicitly by Eqs.\ (\ref{4.5}),
(\ref{4.7}), (\ref{4.11}), and (\ref{4.12}) in a finite domain of
the complex plane $s$ containing the closed interval of the real
axis $-1\leq\text{Re }s\leq 0$.

Now we turn to the calculation of the value of the function
$\zeta^{\text{N}}(s)$ at the point $s=-1$. Integration in Eq.\ (\ref{4.11})
gives
\begin{eqnarray}
 V_0(-1) & = & -2\int_0^{\infty} dy\, y\left\{\ln[-2y
          I_0'(y) K_0'(y)]+\frac{3}{8}t^2\right\}+\frac{13}{16} \nonumber \\
       &=& 2\cdot 0.475215+0.8123=1.76393\,{.} \label{V01}
\end{eqnarray}
From Eqs.\ (\ref{4.7}) and (\ref{3.16}) it follows that
\begin{equation}\label{4.14}
  V_1(-1)=-Z_1(-1)=-\zeta '(-2)=0.03044\,.
\end{equation}
Developing the functions $\zeta (1+s)$ and $\Gamma ((1+s)/2)$ in
Eq.\ (\ref{4.12}) near the point $s=-1$ one obtains
\begin{eqnarray}
 V_2(-1)& = & \frac{1}{8}\left[\zeta(0)+\zeta'(0)(s+1)+1
              +O\left((s+1)^2\right)\right]\cdot
         \left[\frac{2}{1+s}-\gamma+O(s+1)\right] \nonumber \\
        & &  \times \left[3-\frac{13}{4}\, (s+1)\right]
         =-\frac{13}{32}-\frac{3}{16}\gamma +\frac{3}{4}\,\zeta'(0)+
         \left. \frac{3}{8}\frac{1}{s+1}\right|_{s\to -1} {.}
         \label{4.15}
\end{eqnarray}

When calculating $V_3(s)$ for $s=-1$ numerically we cannot use the method applied
in the preceding section because it requires now to take into account the
next terms in the uniform asymptotic expansion (\ref{4.4}).
Instead of this we calculate numerically the first 30 terms in the sum
(\ref{4.9}) with the result\cite{endnote}
\begin{equation}\label{4.16}
  V_3(-1)=-0.04366\,{.}
\end{equation}

Substituting in Eq.\ (\ref{4.9}) the logarithm by its uniform asymptotic
expansion (\ref{4.4}) we derive a rough estimation for $V_3(s)$ without
numerical integration
\begin{eqnarray}
 V_3^{\text{as}}(s)&=&\zeta(3+s)
         \int_{0}^{\infty} dy\, y^{1-s}\frac{d}{dy}\left[t^4
           \left(-\frac{27}{64}+\frac{109}{16}t^2
-\frac{733}{32} t^4+\frac{441}{16} t^6-\frac{707}{64}
           t^8\right) \right] \nonumber \\
&=& \zeta(3+s)
\Gamma\left(\displaystyle{\frac{3-s}{2}}\right)
  \left[\frac{27}{64}\,\Gamma\left(\displaystyle{\frac{3+s}{2}}\right)-
  \frac{109}{32}\,\Gamma\left(\displaystyle{\frac{5+s}{2}}\right)+
  \frac{733}{192}\,\Gamma\left(\displaystyle{\frac{7+s}{2}}\right)\right.
\nonumber \\
 & &
  \left. -\frac{441}{384}\,\Gamma\left(\displaystyle{\frac{9+s}{2}}\right)+
  \frac{707}{7680}\,\Gamma\left(\displaystyle{\frac{11+s}{2}}\right)\right]{.}
\label{4.17}
\end{eqnarray}
For $s=-1$ it gives
\begin{equation}
\label{4.18}
 V_3^{\text{as}}(-1)=-\frac{839}{2^6\cdot 3\cdot 5}\zeta(2)=-\frac{839}{960}
\frac{\pi^2}{6}=-1.43760 \,{,}
\end{equation}
that is  very far from Eq.\ (\ref{4.16}) having only the right sign.

Summing up $V_i$, $i=0, 1, 2, 3$  with allowance for Eq.\
(\ref{3.13}) we arrive at the final result
\begin{eqnarray}
  \zeta^{\text{N}}(-1) &=& -\frac{1}{4\pi
  a^2}\left[\frac{13}{32}+0.95043-\zeta
  '(-2)-\frac{3}{16}\gamma \right . \nonumber \\
&&- \frac{3}{8}\, \ln (2\pi)
   \left. - 0.04366+\left.
  \frac{3}{8}\frac{1}{s+1}\right|_{s\to -1}\right] \nonumber \\
  &=& \frac{1}{a^2}\left(-0.04345-0.0298\left. \frac{1}{s+1}\right|_{s\to
  -1}\right)\,.
 \label{4.19}
\end{eqnarray}

Thus both the zeta functions for Dirichlet and Neumann boundary
conditions have the pole at the point $s=-1$. Hence an additional
renormalization is needed in order for a finite physical value of
the relevant Casimir energies to be obtained.

\section{Vacuum Energy of Electromagnetic Field with Boundary
Conditions on a Semi-Circular Cylinder} Analysis of the spectral
problem for the electromagnetic field with boundary conditions on
a semi-circular cylinder (see Sec.\ II) implies that the zeta
function for this field is the sum of two zeta functions
calculated in the preceding Sections
\begin{equation}\label{5.1}
  \zeta^{\text{EM}}(s)=\zeta^{\text{D}} (s)+\zeta^{\text{N}}(s)\,.
\end{equation}
Substitution of Eqs.\ (\ref{3.19}) and (\ref{4.17}) into
Eq.\ (\ref{5.1}) gives
\begin{eqnarray}
  \zeta^{\text{EM}}(-1)&=&-\frac{1}{4\pi
  a^2}\left[\frac{1}{4}+0.95043-\frac{\gamma}{4}-\frac{1}{4}\,\ln
  (2\pi) - 0.04366+\left.\frac{1}{2}\frac{1}{s+1}
  \right|_{s\to-1}\right] \nonumber \\
  & = &\frac{1}{a^2}\left(
  -0.04401-0.03978 \,\left.\frac{1}{s+1}\right|_{s\to -1}\right)\,. \label{5.2}
\end{eqnarray}

In both the zeta functions $\zeta^{\text{D}}(s)$
and $\zeta^{\text{N}}(s)$ the pole
terms have the same sign. As a result the pole contribution in the
sum (\ref{5.1}) retains. Thus, the situation here proves to be
analogous to that when calculating, in the framework of zeta
technique, the vacuum energy for spheres in spaces of even
dimensions.\cite{Milton,LR,CEK}

As was noted above, we have derived the exact expressions for the
zeta functions in question which determine these functions as
analytic functions of the complex variable $s$ in a finite region
of the plane $s$ containing the closed interval  of the real axis $-1 \leq
\text{Re }s \leq 0$.  It enables one to construct
in a
straightforward way the spectral zeta functions for relevant boundary value
problem on the plane by
making use of the relation \cite{NP}
\begin{equation}\label{5.3}
  \zeta_{\text{s-cir}}(s) = 2\sqrt{\pi}\,\frac{\Gamma
\left(\displaystyle{\frac{s+1}{2}}\right)}
  {\Gamma\left(\displaystyle{\frac{s}{2}}\right)}\,\zeta_{\text{s-cyl}}(s){,}
\end{equation}
where $\zeta_{\text{s-cir}}$ is the Dirichlet or the Neumann zeta
function for a semi-circle, and $\zeta_{\text{s-cyl}}$ is the respective
zeta function for semi-circular cylinder. We shall use this
relation for calculating the values $\zeta_{\text{s-cir}}^{\text{D}}(-1)$ and
$\zeta_{\text{s-cir}}^{\text{N}}(-1)$ which determine the vacuum energy of the
massless scalar fields defined on the half-plane and obeying,
respectively, the Dirichlet or Neumann boundary conditions on a
semi-circle (see Fig.\ 1).

For $\zeta_{\text{s-cir}}^{\text{D}}(-1)$ we get from Eqs.\ (\ref{5.3}).
(\ref{3.4}), and (\ref{3.2})
\begin{equation}\label{5.4}
  \zeta^{\text{D}}_{\text{s-cir}}(-1)=-\frac{1}{\pi a}\sum_{i=1}^{3}Z_i(0)\,.
\end{equation}
When $s=0$ integration in Eq.\ (\ref{3.6}) can be done explicitly with the
result
\begin{equation}\label{5.5}
  Z_1(0)=-\zeta(-1)\int_0^{\infty}dy\ln\frac{y}{\sqrt{1+y^2}}= \frac{1}{12}
\left(-\frac{\pi}{2}
\right )=
  -\frac{\pi}{24}\,.
\end{equation}
From Eq.\ (\ref{3.12}) it follows that
\begin{equation}\label{5.6}
  Z_2(0)=\frac{\pi}{128}\left(\left.\frac{1}{s}\right|_{s\to0}+\gamma\right)\,.
\end{equation}
Numerical integration in Eq.\ (\ref{3.8}) with $s=0$ gives
\begin{equation}\label{5.7}
  Z_3(0)=-0.00304\,.
\end{equation}
Summing up Eqs.\ (\ref{5.5}), (\ref{5.6}), and (\ref{5.7}) we arrive at the result
\begin{eqnarray}
 \zeta^{\text{D}}_{\text{s-cir}}(-1) & =
 &\frac{1}{a}\left(\frac{1}{24}-\frac{\gamma}{128}+
0.00097 -\frac{1}{128}\left.\frac{1}{s}\right|_{s\to 0}\right)
 \nonumber \\
 & =&\frac{1}{a}\left(0.038127-\frac{1}{128}\left.\frac{1}{s}\right|_{s\to 0}
\right){.}
 \label{5.8}
\end{eqnarray}
Following the same way one can write
\begin{equation}\label{5.9}
  \zeta^{\text{N}}_{\text{s-cir}}(-1)=-\frac{1}{\pi a}\sum_{n=0}^{3}V_i(0)\,.
\end{equation}
Using Eq.\ (\ref{4.11}) one gets
\begin{eqnarray}
 V_0(0) &=&
 -\int_0^{\infty} dy \left\{\ln\left[-2yI_0'(y)K_0'(y)\right]+\frac{3}{8}t^2\right\}
 -\frac{\pi}{64} \nonumber \\
 &=& 0.475175-\frac{\pi}{64}\,. \label{5.10}
\end{eqnarray}
From Eqs.\ (\ref{4.7}) and (\ref{5.5}) it follows that
\begin{equation}\label{5.11}
  V_1(0)=-Z_1(0)=\frac{\pi}{24}\,.
\end{equation}
Equation (\ref{4.12}) gives
\begin{equation}\label{5.12}
  V_2(0)=\frac{5\pi}{128}\left(1+\gamma+\left.\frac{1}{s}
\right|_{s\to 0}\right){.}
\end{equation}
For $V_3(0)$ numerical integration in Eq.\ (\ref{4.9}) with $s=0$  gives
\begin{equation}\label{5.13}
  V_3(0)=
-0.005659\,{.}
\end{equation}
Finally, we have
\begin{eqnarray}
 \zeta^{\text{N}}_{\text{s-cir}}(-1) &=&
 \frac{1}{a}\left[-0.15132-\frac{5}{128}\left(\gamma+\frac{5}{3}\right)+
 0.00180
 -\frac{5}{128}\left.\frac{1}{s}\right|_{s\to 0}\right]\nonumber \\
 & =& \frac{1}{a}
 \left(-0.237103-0.0124\left.\frac{1}{s}\right|_{s \to 0}\right)\,.
 \label{5.14}
\end{eqnarray}

Both the functions $\zeta_{\text{s-cir}}^{\text{D}}(s)$
and $\zeta_{\text{s-cir}}^{\text{N}}(s)$
have the pole at the point $s=-1$ with the coefficients of the
same (negative) sign. For electromagnetic field defined on a plane
the boundary conditions reduce to the Neumann conditions. Hence
the relevant zeta function is $\zeta^{\text{N}}_{\text{s-cir}}(s)$.

\section{Conclusion}
In the paper the spectral zeta functions are constructed for
massless scalar fields obeying the Dirichlet and Neumann
boundary conditions on a semi-circular infinite cylinder.
Proceeding from this, the zeta function for electromagnetic field
is also derived for such a configuration. In all three cases, the
final expressions for the relevant Casimir energy contains the
pole contribution. Hence for obtaining the physical result an
additional renormalization is needed.

It is essential that for the zeta functions $\zeta (s)$ the exact
formulas are derived which determine these functions in a finite
region of the complex variable $s$ but not at the vicinity of one
point $s=-1$. This allowed one to get in a straightforward way the
zeta functions for the two dimensional (plane) version of the
boundary value problem at hand, i.e. the zeta functions for scalar
fields defined on a half-plane and obeying the Dirichlet and
Neumann boundary conditions on a semi-circle. In  this case
the final expression for the vacuum energy contains the pole
contributions also.

     Notwithstanding the spectrum of a semi-circular cylinder is very
close to the spectrum of circular one, the zeta function technique
does not give a finite value for vacuum energy in the first case and
does for the second configuration.  In a recent paper\cite{Dowker}
the divergences found in our consideration are attributed to the
existence of edges or corners in the boundaries under investigation.

Closing, it is worth noting that, as far as we know, such boundary
conditions with asymmetric geometry (semi-circular cylinder) has
been considered in the Casimir problem for the first time.
\acknowledgments
Research is supported by the fund MURST ex 40\% and 60\%, art. 65 D.P.R.
382/80.
This work has been accomplished during the visit of V.V.N.\ to the
Salerno University. It is a pleasure for him to thank Professor
G.\ Scarpetta, Drs.\ G.\ Lambiase and A. \ Feoli for warm
hospitality. The financial supports of IIASS and ISTC (Project \# 840)
are acknowledged.
G.L.\ thanks the UE fellowship, P.O.M. 1994/1999, for financial
support.

\begin{figure}  
\caption{The cross section of an infinite semi-circular cylindrical shell of
radius $a$. All the surfaces (bold-faced lines) are assumed
to be perfectly conducting. At the same time this picture presents
the two-dimensional (plane) version of the problem under consideration, i.e.,
the semi-circular boundaries for massless fields defined on the plane.}
\end{figure}

\begin{references}
\bibitem{endnote} The FORTRAN subroutine has been used that approximates
the Bessel functions by Chebyshev's polynomials.
\end{references}
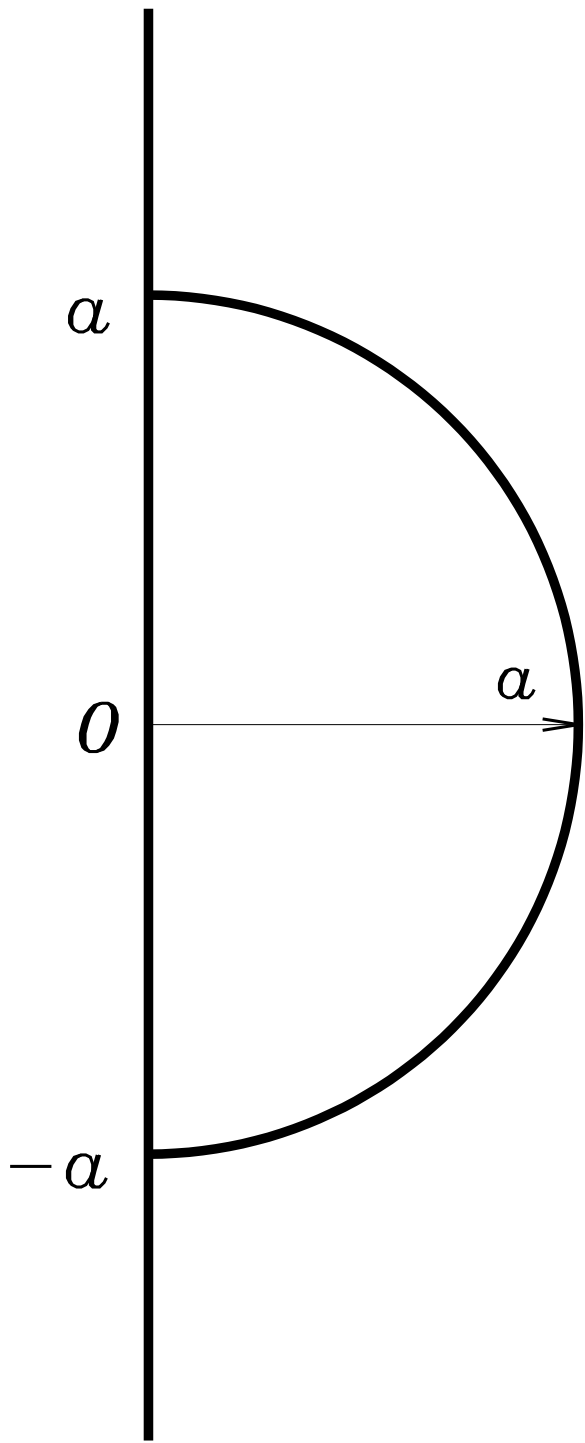
\end{document}